\documentstyle[12pt]{article}    
\begin{document}
\begin{center} {\large \bf   Boundary String Field Theory Approach to High-Temperature Tachyon Potential } \end{center}

\vspace{1cm}
                                                  
\begin{center} { Wung-Hong Huang\\
               Department of Physics\\
               National Cheng Kung University\\
              Tainan,70101,Taiwan\\}\end{center}

\vspace{3cm}
The boundary string field theory approach is used to evaluate the one-loop  tachyon potential.   We first discuss the boundary condition at the two boundaries on annulus diagram and then the exact form of corrected potentials at zero and high temperature are obtained.   The profile of the tachyon potential is found to be temperature independent and tachyon will condense at high temperature.   Our investigations also provide an easy way to prove the universality of the tachyon potential, even if the string is in the thermal background.

\vspace{4cm}
\begin{flushleft}
E-mail:  whhwung@mail.ncku.edu.tw
\end{flushleft}

\newpage

\section{Introduction}

The bosonic open string has tachyonic instability around the usual string vacuum [1-8]. Historically, Kosteleck\'{y} and Samuel [1] studied tachyon condensation in Witten's version of bosonic open string field theory [9] using a level truncation scheme.   In recent,  Sen and Zwiebach [2] had argued that this calculation can describe the decay of a space filling D25-brane to the bosonic string vacuum.  In terms of the D-brane, the existence of the tachyon 
mode in the D-branes system signs that the system is unstable.   It is conjectured that at the minimum of the tachyon potential the D-brane will decay into the closed string vacuum without any D-brane.  In some other unstable systems, such as a non-BPS D-brane or a coincidence D-brane anti-D-brane pair, the tachyon may appear too.    It has also been conjectured (Sen's conjecture [2])that the energy density in some systems will vanish at the minimum tachyon potential  and thus the D-branes disappear .  

  As the tachyon condensation plays an important dynamics in the D-brane systems the tachyon condensation in the string theory has therefore attracted many researchers recently.   In this paper we will investigate the quantum corrected tachyon potential at zero and high temperature.   The later system is the environment in the early universe or densed star, in which the quantum gravity and thus the string theory play a crucial role.   The problem is thus worth studying. 

   As is well known that, at the tree level, the effective tachyon Lagrangian looks like a Higgs field which poses a negative mass squared [1].   Thus, a correct choice of vacuum can stabilize the system and the tachyon field condenses.   In the other words, the tachyon system shows the similar mechanism of spontaneously symmetry breaking in the electroweak model.    However, the temperature is known to be able to restore the tree-level symmetry breaking.    Therefore, it is interesting to see whether the temperature could drastically modified the tachyon potential such that the tachyon field becomes stable and there is no tachyon condensation.  This is the main motivation of this paper.

   We will use the boundary string field theory (BSFT) [10,6,7] to evaluate the one-loop tachyon potential for the bosonic string at zero or high temperature.  Note that the background independent open string field theory proposed by Witten [10] around the year '92--'93  is well suited for the study of tachyon condensation in that only the tachyon field (not the whole string field) acquires the nonvanishing vacuum expectation value.   Therefore we can proceed without any approximation scheme such as the level truncation method [1] to obtain some exact results about the tachyon physics: the tachyon potential and the description of the D-branes as tachyonic solitons.   It is also possible to take a more intuitive approach, generalizing ordinary sigma models [11]. This approach has been used by Kraus and Larsen [12] to investigate the tachyon condensation in the $D{\bar D}$ system.   This is the strategy we pursue in this paper.

    In section II we briefly review the  method of background string field theory (BSFT), and the tree-level tachyon potential calculated by Kraus and Larsen in [12] is reproduced for clear.

    In section III  we first discuss the  boundary condition on the annulus before evaluating the one-loop corrected tachyon potential.   The boundary conditions we adopted are the symmetric boundary condition, which was discussed by Bardakci and Konechny in [13], and diffeomorphism invariant boundary condition, which was discussed by G. Arutyunov, A. Pankiewicz and B. Stefanski  in [13] \footnote {I  will thank the authors of G.A.B [13] who mention the diffeomorphism invariant boundary  condition to me after the first vesion of this paper is posted on arXiv.}.   Then, the boundary conditions are used to evaluated  the quantum corrected potential at zero temperature.    The  one-loop  tachyon potential calculated in the symmetric boundary condition in our approach has the same form as that in [13].   The potential, however, is slightly different from that at the tree level.   It is interesting to see that in the diffeomorphism invariant boundary condition the function form of one-loop tachyon potential is exactly the  same as that at tree level.    However, both results show that the quantum correction does not change the fate of the tachyon condensation.   Therefore, the Sen's conjecture passes the check at the one-loop level.   We also review the paper of CKL [14] in which a more fundamental classical action is defined and is used to study the loop correction in BSFT [14].     Note that our one-loop corrected tachyon potential has the same form as that in [13] but not those in the other literatures [15]. 

   In section IV, after discussing the path integral approach to finite temperature string theory, the high-temperature one-loop corrected tachyon potential is first calculated.  The potential form shows that the temperature does not change the fate of the tachyon condensation.   Thus the Sen's conjecture has also been passed the check at the one-loop level at high temperature.  

   As the tachyon system shows a similar mechanism of spontaneously symmetry breaking in the electroweak model [1],  one may wonder that the temperature effect calculated in section IV does not restore the tree-level symmetry breaking.   Therefore in section V, we use the truncated the string Lagrangian [1] to evaluated the one-loop high-temperature tachyon potential.   The results confirm that the high temperature could not change the fate of the tachyon condensation.   

   In section VI we consider the system including the EM and B field and the property of universality of the tachyon potential in the general background [3], including the thermal environment, are discussed.

 The last section devotes to a short conclusion.
..

\section{Tree-Level Tachyon Potential}
In the boundary string field theory (BSFT) approach to tachyon potential [10] we can begin by computing the partition function of the bosonic string, formally defined as
$$Z(u_{\mu}) = \int  {\cal D}X \, e^{-(S_{{\rm bulk}}+ S_{{\rm bundy}})}~,   \eqno{(2.1)}$$
\noindent
with 
$$S_{{\rm bulk}} = {1 \over 4\pi \alpha'}\int _{\Sigma} d^2 \sigma \sqrt{\gamma}\gamma^{ab}\partial_a X^\mu \partial_b X_\mu~ .\eqno{(2.2)} $$
$$S_{{\rm bndy}} =  {1\over 2\alpha'} \int _{\partial\Sigma} d\theta ~ u_\mu  X^\mu X_\mu~. \eqno{(2.3)}$$
\noindent
Here $\Sigma$ is the world sheet with metric $\gamma$ and coordinates $\sigma$.   $\theta$ is the line coordinate on the boundary ${\partial\Sigma}$.  The tachyon field $T$ is in the quadratic profile, 

$$T=\sum_{\mu} {\frac {u_{\mu}}{2\alpha'}}X_{\mu}^2 , \eqno{(2.4)}$$
\noindent
and BSFT action is given by [10]
$$S(u_{\mu}) = \left( 1+ \sum_{\mu}  {u_{\mu}} - \sum_{\mu}  {u_{\mu}}{\frac{\partial}{\partial u_{\mu}}} \right) Z(u_{\mu}) ..\eqno{(2.5)}$$
\noindent
If we expand the action with respect to $u_{\mu}$ we can then use the relation  (2.4) to rewrite the action in terms of the tachyon field, as that in the field theoretical form.  The tachyon potential is then obtain.   In this section the tree-level tachyon potential calculated by Kraus and Larsen in [12] is reproduced for clear.

   It is known that tree-level world sheet may be conformally mapped into a disc.  Choose the metric on the disc as 
     $$ds^2 = d\rho^2 +\rho^2 d\tau^2 ,~~~\rho\le 1, ~~~  0\le \tau < 2\pi ~,\eqno{(2.6)}$$
\noindent
with the boundary at $\rho =1$, the unique regular solution to the bulk equation of motion,  $\nabla^2 X^\mu=0$, is
$$X^\mu = X_{0}^\mu+\sqrt{\alpha^{\prime}\over 2}\sum_{n=1}^{\infty}
\rho^n\left(X_n^\mu e^{in\tau} + X_{-n}^\mu e^{-in\tau}\right)~. \eqno{(2.7)}$$
\\
\noindent
The bulk action evaluated on this solution is

$$S_{{\rm bulk}} = {1 \over 2} \sum_{n=1}^{\infty} n X_{-n}^\mu X_{n}^\mu~. \eqno{(2.8)}$$
\\
\noindent
The quadratic boundary interaction becomes

$$S_{{\rm bndy}} = {1\over 2\alpha'} \int {d\tau \over 2\pi}  u X^2 |_ {\rho=1} = {1\over 2\alpha'} u X_0^2 + {u\over 2 }\sum_{n=1}^{\infty} X_{-n} X_{n}. ~ \eqno{(2.9)}$$
\\
\noindent
Here the index on $X$ is omitted because we focus on a specific $X$ hereafter for convenience.   We also let $u_{\mu} = u$.   Note that the boundary interaction breaks conformal invariance and therefore takes the theory off-shell, which is the phenomena of tachyon condensation.  

   The partition function of the bosonic string in (2.1) can now be easily evaluated 

$$Z_1(u) = \int {dX_0 \over \sqrt{2 \pi \alpha'}} \prod_{n=1}^\infty {dX_n dX_{-n}\over 2\pi}e^{-(S_{{\rm bulk}} + S_{{\rm bndy}})}= {1 \over \sqrt u} \prod_{n=1}^\infty \left({ 1 \over n + u} \right)~\eqno{(2.10)}$$
\\
\noindent
The zeta function regularization can be used to evaluate the above infinite product, i.e.:

$$\prod_{n=1}^\infty \left( {1 \over n + u}\right) = \exp{\left\{ {d \over ds} \sum_{n=1}^\infty \left(  n + u \right)^{-s}\right\}_{s=0}} = \exp{\left\{ {d \over ds} \left[\zeta(s, u) - u^{-s}\right]\right\}_{s=0}} $$
$$ \exp{\left\{\ln \Gamma(u) - {1 \over 2}\ln 2\pi + \ln u \right\}} = {u ~\Gamma(u) \over \sqrt{2\pi}}~={1 \over \sqrt{2\pi}} ~ (1+O(u)) ,\eqno{(2.11)} $$
\\
\noindent
where the $\zeta(z,q)$ is the Hurwitz zeta function [16].  Therefore the partition function is 

$$Z(u) = {\sqrt{u}~ \Gamma(u) \over  \sqrt{2\pi}}= {1 \over    \sqrt{2\pi}} \left({1\over {\sqrt u}} + O(u)\right). \eqno{(2.12)}. $$
\\
\noindent
The term $O(u)$ in the above equation will contribute to the derivation of tachyon field and not to the tachyon potential, we can thus neglect it.  

   Note that the prefactor ${1 \over \sqrt{2\pi}}$ in (2.12) is that from the value in (2.11) after letting $u=0$.  It is just the partition function without $u$ term. More precisely, it is the partition function with zero value of tachyon.   Therefore the regularized value, i.e. ${1 \over \sqrt{2\pi}}$ shall be absorbed in the renormalize string tension.   Thus the remained extra factor ${1 \over \sqrt{u}}$ will contribute to the tachyon potential. 

  Using (2.5) we can find the action and the tachyon potential at tree level becomes 
$$U(T) = (1+T) e^{-T}, \eqno{(2.13)} $$
up to a renormalized constant. The potential $U(T)$ has two extrema at $T=0$ and $T= \infty$.  $T=0$ corresponds to the original D25-brane, and its energy density $U(0)=1$ (with a renormalized string tension) is exactly equal to the known D25-brane tension. On the other hand, $T=\infty$ is thought of as the `closed string vacuum'.   Because the energy at the new vacuum is vanishing, i.e. $U(\infty)=0$, the Sen's conjecture that the negative energy contribution from the tachyon potential precisely cancels the D25-brane tension is thus proved..   The next subsection is used to investigate the same problem while at the one-loop level.

\section {One-Loop Tachyon Potential}
\subsection {Bulk Action}
 It is known that one-loop world sheet may be conformally mapped into an annulus.  The metric on the annulus is chosen as  
     $$ds^2 = d\rho^2 +\rho^2 d\tau^2 ,~~~a\le \rho\le 1, ~~~  0\le \tau < 2\pi ~,\eqno{(3.1)}$$
\noindent
The boundaries are at $\rho =a, 1$.   The  regular solution to the bulk equation of motion is  

$$X^\mu = X_{0}^\mu+\sqrt{\alpha^{\prime}\over 2}\sum_{n=1}^{\infty}
\left[ \rho^n\left(X_n^\mu e^{in\tau} + X_{-n}^\mu e^{-n\tau}\right) + \rho^{-n} \left( \tilde{X}_n^\mu e^{in\tau} + \tilde{X}_{-n}^\mu e^{-in\tau}\right) \right]. \eqno{(3.2)} $$
\\
\noindent
The bulk action evaluated on this solution is

$$S_{{\rm bulk}} = {1 \over 2} \sum_{n=1}^{\infty} \left[n( 1-a^{2n})  X_{-n}^\mu  X_{n}^\mu~ - n (1-a^{-2n})  \tilde{X}_{-n}^\mu  \tilde{X}_{n}^\mu ~\right]. \eqno{(3.3)}$$

The next work is to evaluate the quadratic boundary interaction.   However, in this stage, there are some controversies on the choice of boundary condition.   The exist literatures on the calculation of annulus-diagram corrected to tachyon potential [13,14] does not use the similar boundary condition and the found corrected potential is different from each other.   So let us first discuss the problems.

\subsection{Boundary Action}

\subsubsection{Symmetry Boundary Condition}
The first boundary action we interesting is that described by

$$S_{{\rm bndy}} = {1\over 2\alpha'} \int {d\tau \over 2\pi}  u X^2 |_ {\rho=1} + {1\over 2\alpha'} \int {d\tau \over 2\pi}  u X^2 |_ {\rho=a}. ~ \eqno{(3.4)}$$
\\
\noindent
In this so called symmetry boundary condition the conditions at the two boundaries, $\rho=1$ and $\rho =a$, are treated on equal footing.   This  seems a simple condition, as discussed by  Bardakci and Konechnyin  [13].  Their discussions are as following.  

    If  we change the annulus coordinate (3.1) to a cylinder with coordinates
$$ \phi = \tau ,~~ t = - {\rm ln} \rho , ~~~~ 0 \le \phi < 2\pi, ~~ - {\rm ln} a \le t \le 0, ~ \eqno{(3.5)}$$
and  represent a string by a boundary state $|S>$, then the partition function which we want to evaluate becomes that with the initial state  $|S>_{t = 0}$ at the right end of the cylinder and the conjugated state $<S|_{t = - {\rm ln}a}$ being the final state at the left end.  The partition function thus have the reflection symmetry about the middle of the cylinder that interchanges the two ends of the cylinder. Thus it seems naturally to choose the symmetry boundary condition, as it treats two boundaries on equal footing.   Not that some papers in [15] does not treat the two boundaries on equal footing, and thus there is relative sign between two terms in (3.4). 

\subsubsection{Diffeomorphism Invariant Boundary Condition}
The second action we interesting is that described by 

$$S_{{\rm bndy}} = {1\over 2\alpha'} ~ u_{\mu}v_{\nu} \int {d\tau \over 2\pi}   X^{\mu}X^{\nu} |_ {\rho=1} + {1\over 2\alpha'} ~ a ~ u_{\mu}v_{\nu} \int {d\tau \over 2\pi}   X^{\mu}X^{\nu}|_ {\rho=a}. ~ \eqno{(3.6)}$$
\\
\noindent
G. Arutyunov, A. Pankiewicz and B. Stefanski [13] first use this boundary condition to evaluated the annulus partition in presence of tachyon for the superstring theory.   In this so called diffeomorphism invariant boundary condition the presence of $a$ in the last term will make the measure of the integral be diffeomorphism invariant.   It is seen that  symmetric conditions easily follow from the diffeomorphism invariant boundary condition by choosing $u_{\mu}=u$, $v_{\nu}= u/\sqrt{a} $.   In the following calculations, however, we shall let $u_{\mu}=v_{\mu}$  as we consider both ends of the string are on the same brane.

\subsection{One-Loop Partition Function and One-Loop Tachyon Potential}

   The boundary action can be written as 

$$S_{{\rm bndy}} = {1 \over 2 \alpha'} \int {d\tau \over 2\pi}  u X^2 |_ {\rho=1} +  c ~ {1\over 2\alpha'} \int {d\tau \over 2\pi}  u X^2 |_ {\rho=a}, ~ \eqno{(3.7)} $$
\\
\noindent
if we take $u_{\mu}= u$.   Thus the above action becomes the symmetry boundary action if $c=1$ and becomes diffeomorphism invariant boundary action if $c=a$.

   Using the solution (3.2) the general boundary interaction (3.7) becomes

$$S_{{\rm bndy}} = {1+c\over 2 \alpha'} u X_0^2 + {u\over 2 }\sum_{n=1}^{\infty}  [ (1+c ~ a^{2n})  X_{-n} X_{n} \hspace{4cm}$$

$$ + (1+ c~a^{-2n})  \tilde{X}_{-n} \tilde{X}_{n} +{1+c\over 2 }\tilde{X}_{n} X_{n} + {1+c\over 2 }\tilde{X}_{-n} X_{n} ]. ~ \eqno{(3.8)}$$
\\
\noindent
Now, the partition function on annulus with the bulk action in (3.3) and boundary action in (3.8) can be easily evaluated to be 

$$Z_1(u) = \int {dX_0 \over \sqrt{2 \pi \alpha'}} \prod_{n=1}^\infty {dX_n dX_{-n}\over 2\pi}e^{-(S_{{\rm bulk}} + S_{{\rm bndy}})} \hspace{2cm}$$
$$=  {1 \over \sqrt {(1+c) u}} \prod_{n=1}^\infty \left[{1 \over { (a^{-n} -a^n)~n^2} } + O(u) \right].~\eqno{(3.9)}$$
\\
\noindent
The term $O(u)$ in the above equation will contribute to the derivation corrections of tachyon field and not to the  tachyon potential, we can thus neglect it. 

\subsubsection {One-Loop Partition Function without tachyon field}
   Note that the infinite product term in (3.9) is just the corresponding terms which will contribute to the partition function but without $u$ term, i.e.  without tachyon field.   Therefore, after the integration over $a$ (from  zero to 1) the divergence value shall be canceled by the counter term to give the finite renormalized string tension.   The exact form is irrelevant to our result and is neglected here. 

\subsubsection {One-Loop Tachyon Potential: Symmetry Boundary Condition}

In the symmetry boundary condition $c=1$, and the remained factor in the partition (3.9) becomes ${1 \over \sqrt{2u}}$.   This factor, as that in (2.12), will contribute to the tachyon potential.  Now, the infinite product term in (3.9) is the corresponding terms discussed in the above subsection and the one-loop tachyon potential now becomes
 
 $$U(T) = (1+2T) e^{-2T},  \eqno{(3.10)} $$
\noindent
up to a renormalized constant term which depends on the string tension.   The above one-loop tachyon potential  agrees with that evaluated by Bardakci and Konechny in different approach [13].  

\subsubsection {One-Loop Tachyon Potential: Diffeomorphism Invariant Boundary Condition}

In the diffeomorphism invariant boundary condition $c=a$, and the remained factor in the partition (3.9) becomes ${1 \over \sqrt{(1+a)u}}$.  This factor depends on the internal radius $a$ and shall be included into the infinite product term in (3.9) before taking an integration over $a$ (from  zero to 1) to find the partition function.    To handle this factor we see that the divergence is coming from the region $a \to 0 $ during integrating over $a$, therefore we may take an approximation by  ${1 \over \sqrt{(1+a)u}} \approx  {1 \over \sqrt{u}}$.  Now this factor is exactly that in the tree level, i.e. (2.12), and the one-loop tachyon potential becomes

 $$U(T) = (1+T) e^{-T},  \eqno{(3.11)} $$
\noindent
up to a renormalized constant term which depends on the string tension.   

\subsection {CKL Approach to the Loop Corrected Tachyon Condensation}

In this subsection we review the paper of Crup, Kraus and Larsen (CKL) [14] in which a more fundamental classical action is defined and used to study the loop correction in boundary string field theory (BSFT).     

   BSFT has so far provided a good understanding of tachyon condensation at the classical level. The question addressed in CKL paper is how to include quantum corrections in BSFT.   It is known that conformal invariance is broken by the tachyon background and one is left with an ambiguity as to the choice of Weyl factor of the world-sheets. This ambiguity did not arise at the level of the disk because the definition of BSFT demands that the Weyl factor be rotationally invariant, and the remaining freedom can be compensated for by a redefinition of couplings.  

   Crup, Kraus and Larsen had found that to develop a consistent formalism in which gravity does couple to the energy momentum tensor the two boundaries on the annulus (or cylinder) should be on an equal footing.   They also found some discrepancies which, however, only arise for nonconstant tachyon backgrounds.  

   According to the CKL comment,  if we used the diffeomorphism invariant boundary condition (3.6) we will encounter the difficulties such as the breakdown of the Fishler-Susskin mechanism.    Beside, the conformal invariance will be broken in the system including the boundary interaction, but we could still expect that the system is invariant under a map which mapping the inner boundary  of annual onto the outer boundary.  The diffeomorphism invariant boundary condition (3.6) does not respect this symmetry while the symmetry boundary condition (3.4) does.   Therefore the one-loop potential in (3.10) is a more reliable result. 

   Note that Crup, Kraus and Larsen had made a farther step.   They find that the correct procedure is to use the integrated vertex operator to ensure the full set of the string gauge invariances. They use the totally novel action to find the loop corrected tachyon potential.   Their result is 

 $$U(T) = [(1+T) e^{-T}]^2,  \eqno{(3.12)} $$
\noindent
up to a renormalized constant term.    

   The quantum corrected potential $U(T)$ , (3.10) or (3.11), has two extrema at $T=0$ and $T= \infty$ as that in the tree level.  The tachyon will roll down towards the vacuum at $T=\infty$ in which the new vacuum has vanishing energy.   This proves the Sen's conjecture at one-loop level. 

   The quantum corrected potential $U(T)$ in (3.12) has three extrema at $T=-1$, $T=0$ and $T= \infty$.   The tachyon will roll down towards the vacuum, at $T=-1$ or $T=\infty$,  both of which also have vanishing energy.   This proves the Sen's conjecture at one-loop level.

   Note that the existence of two vacua in the loop corrected potential (3.12) is a new phenomena and deserves to be investigated furthermore.   It also remains to see whether a field redefinition can transfer (3.10) to (3.12). 


\section{Tachyon Potential: Finite Temperature}

    In the above section we use path integral approach to evaluate the stringy partition function.    As well known, the path integral approach can be easily extended to investigate the finite temperature string theory [17].   In this approach the Euclidean spacetime with time $X^0$ will be compactified on a circle of circumference.   Temperature will affect the string gas as the string
can wrap around the compact time direction with a given winding number $\ell$, i.e. $X^0(\rho,\tau +2\pi)=X^0(\rho,\tau)+ \ell \beta$.  This will affect  the zero modes of the bosonic string embedding field $X^0$ and thus can be incorporated by adding a term ${n\beta \over 2\pi} \tau$ to its mode expansion.  

   Note that in string perturbation theory, the disc amplitude is unmodified at finite temperature, because the disc worldsheet cannot wrap the cylindrical target space and so cannot distinguish between a compactified and an uncompactified spacetime.  Thus there is not temperature effect at tree level and the first corrections due to temperature will appear in the annulus amplitude.   This is the work of this section.

   From the above discussions we see that at temperature $1/\beta$ the regular solution to the bulk equation of motion is  
$$X^\mu (\beta)= X^\mu + \frac{\ell \beta}{2\pi} \tau ~ \delta _{\mu,0}, \eqno{(4.1)} $$
\noindent
in which $X^\mu$ is the solution in (3.2).  The bulk action evaluated on this solution is
$$S_{{\rm bulk}}(\beta) = S_{{\rm bulk}} + {\ell^2\beta^2\over 16\pi}, \eqno{(4.2)}$$
\noindent
in which $S_{{\rm bulk}}$ is calculated in (3.3).  Using the symmetry boundary condition the quadratic boundary interaction becomes
$$S_{{\rm bndy}}(\beta) = S_{{\rm bndy}} + (1+c) ~u {\ell^2\beta^2\over 24}, ~ \eqno{(4.3)}$$
\noindent
in which $S_{{\rm bndy}}$   is evaluated in (3.8).

   The finite temperature partition function on annulus can be evaluated as before, except that the winding modes $\ell$ shall be summed.   At high temperature we can use the Poisson summation formula to perform the summation, i.e.

 $$ \sum_{\ell =-\infty}^{\infty} exp\left[ - \left({1\over 16 \pi} + {(1+c) u\over 24}\right) \ell^2\beta^2\right]  = \int _{-\infty}^ {\infty} dt ~ exp\left[ - \left({1\over 16 \pi}  + {(1+c)u\over 24}\right) \beta^2  t^2 \right] +$$
$$ 2  \sum_{n=1}^{\infty} \int _{-\infty}^ {\infty} dt ~ exp\left[ - \left({1\over 16 \pi}  + {(1+c)u\over 24}\right) \beta^2  t^2 \right] cos(2 n \pi t) $$
$$ = {\sqrt \pi \over \sqrt { {1\over 16 \pi} + {(1+c)u\over 24}} \beta} \left[ 1+ 2  \sum_{n=1}^{\infty} exp\left(- {n^2\pi^2\beta^{-2}\over {1\over 16 \pi}  + {(1+c)u\over 24}}  \right)  \right]$$

$$\approx {4\pi\over\beta}[1 + O(u)],  \hspace{6cm}   \eqno{(4.4)}$$
\\
\noindent
which is a good approximation at high temperature.  Note that as discussed before, the term $O(u)$ does not contribute to tachyon potential and thus is neglected.  

   From (4.4) and result in section 3.3 we finally obtain the high temperature one-loop corrected tachyon potential 

$$U(T,\beta) ={4\pi\over\beta} ~ U(T),  \eqno{(4.5)} $$
\noindent
in which $U(T)$ is the zero-temperature one-loop tachyon potential evaluated in (3.10) and (3.11).  Thus we see that, likes that at zero temperature, the tachyon at high temperature will also roll down towards the vacuum at $T=\infty$ in which the new vacuum has vanishing energy.   This prove the Sen's conjecture at high temperature.


\section {One-Loop Tachyon Potential in Level Truncation Scheme}

As discussed in reference [1] the  tachyon system shows a similar mechanism of spontaneously symmetry breaking in the electroweak model,  one may wonder that the temperature effect does not restore the tree-level symmetry breaking.   Although it is not easy to give a simple explanation we would like to give the following  observation.   
  
   In the truncated approximation the string Lagrangian determined to the level (1.2) is [1] 
 
$$L_{(0,0)}= -{1\over 2}\partial_{\mu}\phi  \partial^{\mu}\phi +{1\over 2} \phi^2-{1\over 3} \left(\frac{3\sqrt{3}}{4} \right)^3\tilde{\phi}^3 - {1\over 2}\partial_{\mu} A_{\nu} \partial^{\mu}A^{\nu} - \frac{3\sqrt{3}}{4} \tilde{\phi}\tilde{A_{\mu}}\tilde{A^{\mu}} $$
$$-\frac{3\sqrt{3}}{8} \left(\partial_{\mu} \partial_{\nu} \tilde{\phi} \tilde{A^{\mu}} \tilde{A^{\nu}} + \tilde{\phi} \partial_{\mu} \tilde{A^{\nu}}\partial_{\nu}\tilde{A^{\mu}} - 2 \partial_{\mu} \tilde{\phi} \partial_{\nu} \tilde{A^{\mu}} \tilde{A^{\nu}} \right),  \eqno{(5.1)}$$
\noindent
in which 
$$\tilde{\phi}(x) \equiv exp\left(-\ln\frac{4}{3\sqrt{3}}\partial_{\mu} \partial^{\mu}\right)\phi(x) , ~~~ \tilde A_{\lambda}(x)\equiv exp \left(-\ln\frac{4}{3\sqrt{3}}\partial_{\mu} \partial^{\mu}\right)  A_{\lambda}(x)  \eqno{(5.2)}$$
\noindent
We let $\alpha'=1$ for convenience in the above relation.

    The positive sign before  $\phi^2$ means that the symmetry is broken at the tree level.   The higher derivative terms in tachyon field $\phi$ are known to play important roles in determining the spectrum at the nonperturbative vacuum, but, as the first approximation, we will simply set $\tilde{\phi}=\phi$.  This assumption is true if $\phi(x)$ does not intensively fluctuate.   The Lagrangian then become $\phi^3$ theory coupled to EM field in the conventional  field theory.   Using the method of paper [17] we can find the high-temperature effective potential 

$$V_{eff} = C { 1\over \beta^2} \left[-1 +  2 \left(\frac{3\sqrt{3}}{4} \right)^3 \phi\right],  \eqno{(5.3)}  $$
\noindent 
in which C is a positive number which is dependent on the spacetime dimensions.    As is easily seen, this corrected potential could not stabilize the field at $\phi=0$ and thus could not restore the symmetry breaking at tree level.  

   In fact, one can prove a general property that the temperature can restore the symmetry breaking in $\phi^4$ theory but cannot restore the symmetry breaking in $\phi^3$ theory,    Note that the result of (5.3) is very rough as it considers only the lower truncated string Lagrangian determined at the level (1,2).  But we thought it is still interesting. 


\section{Universality of Tachyon Potential:  String with EM and B Field}

The above method could be extended to investigate the tachyon condensation in B and F field.  In this case the bulk action is 

    $$S_{\rm bulk}(B,F) = S_{\rm bulk} +  {1 \over 4\pi \alpha'}\int _{\Sigma} d^2 \sigma ~ \epsilon^{ab} (B_{\mu\nu} - F_{\mu\nu}) \partial_a X^\mu \partial_b X^\nu~ .\eqno{(6.1)} $$\\
\noindent
in which $S_{\rm bulk}$ is defined in (2.2).   The second term in (6.1) does not change the tachyon potential.   This can be seen from the following observation.

   From the sections II - IV we see that only the first term in $S_{\rm bndy}$, i.e. $\sim u  X_0^2$ in (2.9) or (3.8), will itself contribute to the tachyon potential.  The remaining $u$-dependent terms in $S_{\rm bndy}$ and $S_{\rm bulk}$ will contribute to the kinetic term or higher derivations of tachyon field, which is irrelevant to the tachyon potential and may be neglected.  Thus, when the bulk action does not depend on  $X_0^{\mu}$ (which is defined in (2.7)) then the path integration over the action can only affect the renormalized string tension, as the case in section III, and, at most, contribute a prefactor in the potential form, as the case in section IV.   Therefore, no matter what the backgrounds may be, if the bulk action does not depend on the mode $X_0^{\mu}$, then the backgrounds can only affect the renormalized string tension and, at most,  contribute a prefactor in the potential form.   The second term in (6.1) is proportional to $\partial_a X^\mu \partial_b X^\nu~ $  and does not depend on the zero model  $X_0^{\mu}$, thus it will not correct the potential, even the system is in a heat bath.   This leads us to conclude that the tachyon potential is universal [3], in the sense that the potential does not depend on the background which string is coupled to or living in. 

\section{Discussions}

     Tachyon condensation is a conceptually simple process of  fields rolling down a potential towards a  minimum.   Sen had conjectured that the negative energy contribution from the tachyon potential at the minimum point will precisely cancels the D25-brane tension [2].  This conjecture has been proved at the tree level [1-6] and one-loop level [13-15].  

  In this paper we use a intuitive approach of boundary string field theory, which has been used by Kraus and Larsen [12], to evaluate the one-loop tachyon potential of the bosonic string.   

   We first use the symmetry boundary condition to evaluate the one-loop tachyon potential.  In this boundary condition our results  agree with that calculated by Bardakci and Konechny [13] who use a different approach.  

   We next use the diffeomorphism invariant boundary condition [13] to evaluate the one-loop tachyon potential.  We find that in this seems more natural condition the quantum corrected form of the tachyon potential is exactly as that in the tree level.

    We also review the paper of CKL [14].   According to the CKL comment,  if we used the diffeomorphism invariant boundary condition (3.6) we will encounter the difficulties such as the breakdown of the Fishler-Susskin mechanism .   Thus the one-loop potential in (3.10) is a more reliable result.   As CKL adopt a novel fundamental classical action to study the loop correction in BSFT their result is different from others.    It remains to investigate the implications of their results.

   As the string at high temperature is that in the early universe or densed star, the investigation about the tachyon condensation at high temperature is  physically interesting.  We extend the approach to the finite temperature.  The high-temperature tachyon potential calculated in this paper is found to have the same form as that in zero temperature, up to a temperature dependent string tension.  The Sen's conjecture thus passes the check at the one-loop level at high temperature.       

     As the  tachyon system shows the similar mechanism of spontaneously symmetry breaking in the electroweak model,  one may wonder that the temperature effect does not restore the tree-level symmetry breaking.    We thus use the truncated string Lagrangian [1] to evaluate the one-loop high-temperature  tachyon potential, as that in the particle field theory [18].   Our results confirm the fact that the high temperature could not change the fate of the tachyon condensation.   

    We have also considered the theory including the EM and  B field.   Then we find that our method could also provide an easy way to prove the universality of the tachyon potential, even if the string is in a thermal background.

   Finally we want to mention that the D-brane -anti-D-brane system at finite temperature has been investigated in [19].  It was found that a finite temperature leads to the reappearance of open string degrees of freedom and that at a sufficiently large temperature the open string vacuum becomes stable.   Thus the tachyon field at the open string vacuum is no longer tachyonic, and so the $D{\bar D}$ system is stable.   We will use the prescription of this paper to study the finite temperature $D{\bar D}$ system in the next paper. 

\newpage
\begin{enumerate}
\item  K. Bardakci, " Dual Models and Spontaneous Symmetry Breaking", Nucl.Phys. B68(1974)331;\\
K.Bardakci and M.B.Halpern, "Explicit Spontaneous Breakdown in a Dual
Model", Phys.Rev.D10 (1974)4230;\\
K.Bardakci and M.B.Halpern, "Explicit Spontaneous Breakdown in a Dual
Model II: N Point Functions", Nucl. Phys. B96(1975)285;\\
K.Bardakci, "Spontaneous Symmetry Breakdown in the Standard Dual String
Model", Nucl.Phys.B133(1978)297,\\
V.A. Kosteleck\'{y} and S. Samuel,  ``On a Nonperturbative Vacuum for the Open Bosonic  String," {Nucl. Phys.} {\bf B336} (1990) 263.
\item  A. Sen, ``Tachyon Condensation on the brane antibrane system,'' JHEP {\bf 9808} (1998) 012, hep-th/9805170,\\
A. Sen and B. Zwiebach, ``Tachyon Condensation in String Field Theory,'' JHEP {\bf 0003} (2000) 002, hep-th/9912249.

\item A. Sen, ``Descent Relations Among Bosonic D-branes,'' Int.J. Mod.
Phys. {\bf A14} (1999) 4061, hep-th/9902105,\\
A.~Sen, ``Universality of the Tachyon Potential ,'' JHEP {\bf 9912} (1999) 027, hep-th/9911116. 

\item  W. Taylor,  ``D-brane effective Field Theory from  String Field Theory,'' Nucl. Phys. {\bf 585} (2000) 171, hep-th/0001210;\\
N. Moeller and W. Taylor,  ``Level Truncation and the Tachyon in Open Bosonic String Theory,'' Nucl. Phys. {\bf 583} (2000) 105, hep-th/0002237.

\item 1.  R. de Mello Koch, A. Jevicki, M. Mihailescu and R. Tatar, ``Lumps and p-Branes in Open String Field Theory", Phys. Lett. {\bf B482} (2000) 249, hep-th/0003031;\\
N. Moeller, A. Sen and B. Zwiebach, ``D-branes as Tachyon 
Lumps in String Field Theory," JHEP {\bf 0008} (2000) 039, hep-th/0005036;\\
R. de Mello Koch and J.P. Rodrigues, ``Lumps in Level truncated open string field theory", Phys. Lett. {\bf B495} (2000) 237, hep-th/0008053;\\
N. Moeller, ``Codimension two lump solutions in string field 
 theory and tachyonic theories," hep-th/0008101.

 \item  A.A. Gerasimov and S.L. Shatashvili, ``On Exact Tachyon Potential in Open String Field Theory," JHEP {\bf 0010} (2000) 034, hep-th/0009103;\\
D. Kutasov, M. Mari\~{n}o and G. Moore, ``Some Exact Results on Tachyon Condensation in String Field Theory," JHEP{\bf 0010} (2000) 045, hep-th/0009148;\\
J. R. David, ``Tachyon condensation in the D0/D4 system," hep-th/0007235.
\item D. Ghoshal and  a. Sen, ``Normalization of the background independent open string field theory action'',  JHEP {\bf 0011} (2000) 0214, hep-th/0009191.
\item K. Ohmori, ``A Review on Tachyon Condensation in Open String Field Theories'', hep-th/0102085.

\item E. Witten, ``Noncommutative Geometry and String Field
Theory,'' Nucl. Phys. {\bf B268} (1986) 253.

\item E. Witten,``On background independent open string field theory'', Phys. Rev. {\bf D46} (1992) 5467, hep-th/9208027;\\
E.Witten,``Some computations in background independent off-shell string theory'', Phys. Rev.  {\bf D47} (1993) 3405, hep-th/9210065;\\
S. L. Shatashvili, ``Comment on the background independent open string theory'', Phys. Lett. {\bf B311} (1993) 83, hep-th/9303143; ``On the problems with background independence in string theory'', hep-th/9311177.

\item E. S. Fradkin and A. A.~Tseytlin,``Nonlinear Electrodynamics From Quantized Strings,'' Phys.  Lett.  {\bf B163} (1985) 123;\\
O. D. Andreev and A. A. Tseytlin, ``Partition Function Representation For The Open Superstring Effective Action: Cancelation Of M\"{o}bius Infinities And Derivative Corrections To Born-Infeld Lagrangian'', Nucl. Phys. {\bf B311} 
(1988) 205.

\item P. Kraus and F. Larsen, ``Boundary String Field Theory of the  $D{\bar D}$'', Phys. Rev.  {\bf D63} (2001) 1060004, hep-th/0012198;\\
T.Takayanagi, S.Terashima and T.Uesugi,
"Brane-Antibrane Action from Boundary String Field Theory", JHEP {\bf 0103} (2001) 019, hep-th/0012210.

\item K. Bardakci and A. Konechny, ``Tachyon Condensation in Boundary String Field Theory at one loop'', to appear in  Nucl. Phys., hep-th/0105098 \\
G. Arutyunov, A. Pankiewicz and B. Stefanski, ``Boundary superstring field theory annulus partition function in the presence of tachyons,'' JHEP {\bf 0106} (2001) 049, hep-th/0105238.

\item B. Craps,  P. Kraus and F. Larsen, ``Loop corrected Tachyon condensation '',  JHEP {\bf 0106} (2001) 062, hep-th/0105227.

\item  R. Rashkov, K. S. Viswanathan and Y. Yang, ``Background independent open string field theory with constant B field on the annulus'', hep-th/0101207;\\
O. Abdreev, ``Some calculations of partition functions and tachyon potentials in background independent off-shell string theory'', Nucl. Phys. {\bf B598} (2001) 151, hep-th/0101218;\\
T. Suyama, ``Tachyon Condensation and Spectrum of String on D-banes'', hep-th/0102192;\\
K. S. Viswanathan and Y. Yang, ``Tachyon Condensation and background independent superstring field theory '', hep-th/0104099;\\
M. Alishahiha, ``One-loop correction of the tachyon action  in boundary superstring field theory'', hep-th/0104164.

\item E. Elizalde, S. D. Odintsov, A. Romeo, A. A. Bytsenko, and Zerbini, {\it Zeta Regularization Techniques with Applications},World Scientific, (1994).

\item  J. Ambjorn, Y. M. Makeenko, G. W. semenoff and R. J. Szabo, ``String theory in electromagnetic fields'',  hep-th/0012092.

\item L. Dolan and R. Jackiw, ``Symmetry breaking at finite temperature'',   Phys. Rev. {\bf D 9} (1974)3320.

\item Ulf H. Danielsson, Alberto Guijosa, Martin Kruczenski, ``Brane-Antibrane Systems at Finite Temperature and the Entropy of Black Branes
  '',  hep-th/0106201.

\end{enumerate}
\end{document}